\divide\pdfpxdimen by 10

\documentclass[a4paper,twocolumn,11pt]{quantumarticle}
\pdfoutput=1
\usepackage[utf8]{inputenc}
\usepackage[english]{babel}
\usepackage[T1]{fontenc}
\usepackage{etex}
\usepackage{amsmath,amssymb,amsthm}
\usepackage{hyperref}
\hypersetup{colorlinks=true,citecolor=blue,urlcolor=blue}
\usepackage[pdftex]{graphicx}
\usepackage{times,txfonts}
\usepackage{braket}
\usepackage{color}
\usepackage{natbib}
\usepackage{amsmath,blkarray}
\usepackage{mathtools}
\usepackage[normalem]{ulem}
\usepackage{latexsym}
\usepackage{tabularx, booktabs}
\usepackage{graphics,epstopdf}
\usepackage{graphicx}
\usepackage{float}
\usepackage{amsfonts}
\usepackage{tikz}
\usetikzlibrary{quantikz2}
\usepackage{color,soul}
\usepackage{algorithm}
\usepackage{comment}
\usepackage{algpseudocode}
\usepackage{bm}
\algnewcommand\algorithmicforeach{\textbf{for each}}
\algdef{S}[FOR]{ForEach}[1]{\algorithmicforeach\ #1\ \algorithmicdo}

\usepackage{multirow}
\usepackage{appendix}
\usepackage{url}

\usepackage[normalem]{ulem}
\useunder{\uline}{\ul}{}
\usepackage{bbold}
\setcitestyle{square,numbers,comma}
\usepackage{ragged2e}

\begin{document}
\title{Symmetry-enhanced Counterdiabatic Quantum Algorithm for Qudits}

\author{Alberto Bottarelli}\email{alberto.bottarelli@unitn.it}\affiliation{Pitaevskii BEC Center, CNR-INO and Dipartimento di Fisica, Universita di Trento, I-38123 Trento, Italy}\affiliation{INFN-TIFPA, Trento Institute for Fundamental Physics and Applications, Trento, Italy}\orcid{0000-0002-5056-4109}
\author{Mikel Garcia de Andoin}\affiliation{TECNALIA, Basque Research and Technology Alliance (BRTA), 48160 Derio, Spain}\affiliation{Department of Physical Chemistry, University of the Basque Country UPV/EHU, 48940 Leioa, Spain}\affiliation{EHU Quantum Center, University of the Basque Country UPV/EHU, 48940 Leioa, Spain}\orcid{0000-0002-5009-7109}
\author{Pranav Chandarana}\affiliation{Department of Physical Chemistry, University of the Basque Country UPV/EHU, 48940 Leioa, Spain}\affiliation{EHU Quantum Center, University of the Basque Country UPV/EHU, 48940 Leioa, Spain}\orcid{0000-0002-3890-1862}
\author{Koushik Paul}\email{koushikpal09@gmail.com}\affiliation{Department of Physical Chemistry, University of the Basque Country UPV/EHU, 48940 Leioa, Spain}\affiliation{EHU Quantum Center, University of the Basque Country UPV/EHU, 48940 Leioa, Spain}\orcid{0000-0002-2732-9629}
\author{Xi Chen}\affiliation{Instituto de Ciencia de Materiales de Madrid (CSIC), Cantoblanco, E-28049 Madrid, Spain}\orcid{0000-0003-4221-4288}
\author{Mikel Sanz}\affiliation{Department of Physical Chemistry, University of the Basque Country UPV/EHU, 48940 Leioa, Spain}\affiliation{EHU Quantum Center, University of the Basque Country UPV/EHU, 48940 Leioa, Spain}\affiliation{IKERBASQUE, Basque Foundation for Science, 48009 Bilbao, Spain}\affiliation{Basque Center for Applied Mathematics BCAM, 48009 Bilbao, Spain}\orcid{0000-0003-1615-9035}
\author{Philipp Hauke}\affiliation{Pitaevskii BEC Center, CNR-INO and Dipartimento di Fisica, Universita di Trento, I-38123 Trento, Italy}\affiliation{INFN-TIFPA, Trento Institute for Fundamental Physics and Applications, Trento, Italy}\orcid{0000-0002-0414-1754}

\begin{abstract}
Qubit-based variational quantum algorithms have undergone rapid development in recent years but still face several challenges. In this context, we propose a symmetry-enhanced digitized counterdiabatic quantum algorithm utilizing qudits instead of qubits. This approach offers three types of compression compared to conventional variational circuits. First, compression in the circuit depth is achieved by counterdiabatic protocols. Second, number of particles is reduced by replacing qubits with qudits.
Lastly, the number of parameters is reduced by employing the system's symmetries. We illustrate this approach by tackling a graph-based optimization problem \textsc{Max}-$3$-\textsc{Cut} and a highly-entangled state preparation, the qutrit $W$ state. As our numerical results show, we achieve a better convergence with a lower circuit depth and less measurement overhead in all the cases considered. This work leads to a better design of shallow variational quantum circuits, improving the feasibility of their implementation on near-term qudit devices.
\end{abstract}
\maketitle

\section{Introduction}
Over the past decade, quantum computing has made significant strides, showing great promise for advancing both scientific research and industrial applications~\cite{Preskill2018NISQ, Bravyi2020, beverland2022assessing}. This paves the way for solving complex computational problems, previously considered insurmountable using classical computers~\cite{arute2019quantum,madsen_quantum_2022,kim2023evidence,JianWeiPan2020}. Quantum algorithms like Grover's search~\cite{grover1996fast} and Shor's factorization~\cite{shor1999polynomial} have been proposed to attain quantum advantage, particularly in a fault-tolerant quantum computing scenario. However, achieving practical quantum utility remains challenging in the current noisy intermediate scale quantum (NISQ)~\cite{Preskill2018NISQ, Kishor2022} era, despite several attempts~\cite{kim2023evidence, king2024computationalsupremacy}. This is due to limiting factors like decoherence, dephasing, and large error rates in the present quantum hardware. Most of the current quantum computation is dominated by a qubit-based model, developed to imitate the binary model of classical computation. 
This might not necessarily be optimal for maximizing computational efficiency and exploiting the full potential of quantum hardware capabilities.

In recent years,  there has been a growing interest in utilizing quantum systems with multiple levels, known as qudits, as fundamental components in quantum computing. Qudits offer the potential to enhance computational power and efficiency by reducing the number of required computational units~\cite{wang2020qudits}.
A qudit is formally defined as a quantum system with $d>2$ levels, where the local dimension $d$ varies depending on the physical nature of the system. 
Thanks to advances in quantum hardware and control techniques, universal qudit quantum computers have been realized with trapped $^{40}Ca^+$ ions~\cite{ringbauer2022universal}, superconducting circuits~\cite{fischer2023universal}, Rydberg atoms~\cite{gonzalez2022hardware} and photonic devices~\cite{chi2022programmable}.

Qudit-based quantum computing presents many advantages compared to the usual $d=2$ counterpart~\cite{gao2023role}. Representing a given Hilbert space on a qudit quantum computer requires fewer particles by a factor of $\log_d2$, than a qubit-based quantum computer. This is particularly favorable for NISQ devices since the increased number of qubits needed to represent a large Hilbert space becomes more vulnerable to noise and more challenging to control~\cite{Bravy2022future, osman2023}. The high dimensionality of the qudits also makes them exploitable in different areas of quantum computing. For instance, in quantum simulation, they can be utilized to investigate high-order truncation of quantum link models~\cite{Chandrasekharan:1996ih,Popov2024}. Moreover, many optimization problems like graph coloring and \textsc{Max}-$k$-\textsc{Cut}, which are not based on binary variables, are naturally suited to qudit platforms. Both of these tasks can be studied either through qudit circuit-based variational quantum algorithms (VQA) ~\cite{Popov2024,deller2023quantum} or qudit-based adiabatic computing~\cite{amin2013adiabatic}.

VQAs offer a promising approach to bridge the gap between current quantum capabilities and the goal of achieving quantum utility with the support of classical algorithms~\cite{cerezo2021variational}. For instance, the quantum approximate optimization algorithm (QAOA)~\cite{farhi2014quantum}, derived from the principles of adiabatic quantum computing, has been proposed to address a wide range of problems, including combinatorial optimization~\cite{blekos2024review}.
However, these algorithms require substantial circuit depth to attain the desired solution, especially for large system sizes. The quality of solutions obtained from VQAs can be improved through various approaches like efficient parameterization and improved initialization strategies~\cite{volkoff2021large,grant2019initialization}. { Apart from these, quantum control methods such as }Counterdiabatic (CD) protocols~\cite{del2013shortcuts,chen2010shortcut} can also be utilized to design shallow quantum circuits that reduce the resource requirement and obtain solutions~\cite{YaoLinBukov21}.

In this article, we { propose} a symmetry-enhanced CD-inspired variational ansatz for qudit systems. In Ref.~\cite{Sauvage2024}, symmetries of graph-defined problems were used to improve the performance of VQAs and reduce the overhead on both the classical and quantum hardware. Utilizing symmetries is crucial because they reduce the algorithm's resources in the form of the number of parameters, which is a bottleneck for current VQAs. In this ansatz, the symmetries of the system are exploited to significantly reduce the number of free optimizable parameters. We define the theoretical framework that merges digitized-counterdiabatic quantum optimization with graph permutation symmetries to achieve compression in three regimes. First, information compression is achieved by using qudits instead of qubits. Circuit compression is achieved using CD protocols, and finally compression in parameter space by leveraging symmetries. We demonstrate the efficiency of our method by implementing it on the \textsc{Max}-$k$-\textsc{Cut} graph problem and $W$ state preparation to show that this combination significantly compresses the variational protocols.

This article is organized as follows: In Sec.~\ref{sec: preliminaries}, we review qudit-based quantum computing on trapped ions, variational algorithms, and CD protocols. In Sec.~\ref{sec: Theory symmetric}, we derive the theory of symmetric CD terms. In Sec.~\ref{sec: benchmark}, we numerically test the proposed symmetric CD ans\"{a}tze and compare it to their fully parametrized counterparts and QAOA. Finally, we provide an overview of our results and conclude in Sec.~\ref{sec: conclusion}.
\section{Preliminaries} \label{sec: preliminaries}
Propelled by recent advances in their experimental control \cite{ringbauer2022universal,fischer2023universal,gonzalez2022hardware,chi2022programmable}, qudit-based algorithms and their applications in solving diverse problems are currently receiving significant interest in the quantum computing community~\cite{wang2020qudits}. This section provides an overview of the foundational elements of qudits, which outlines the definition of qudit operations crucial for the present study. Additionally, it explores their role in designing qudit-based VQAs, highlighting their potential for efficient problem-solving. Furthermore, the section briefly discusses the design of CD ans\"{a}tze, an advanced technique aimed at enhancing VQAs.
\subsection{Qudits gates and basis}
A qudit is defined as a quantum system with $d$ possible states and local Hilbert space with dimension $\mathcal{C}^d$, with the local computational basis identified as ${\ket{0},...,\ket{d-1}}$. Unitaries acting on the qudit are part of the $U(d)$ unitary group and can be implemented with arbitrary error $\epsilon$ via a universal gate set. Just like the qubit case, this set will depend on the physical realization of the qudit. { For} example, qudits can be realized by splitting the $4^2S_{\frac{1}{2}}$ and $3^2D_{\frac{3}{2}}$ energy levels of trapped $^{40}\textrm{Ca}^+$ ions with a magnetic field, which results in a total of $8$ levels for the qudit~\cite{ringbauer2022universal}. The universal gate set for such a system is provided by equatorial rotations 
\begin{equation}\label{eq: rotation}
    R^{i,j}(\vartheta,\phi) =\exp(-i\theta\sigma^{i,j}_\phi/2),
\end{equation} 
combined with the entangling MS gate
\begin{equation}\label{eq: MS}
    \textrm{MS}^{i,j}(\vartheta,\phi) = \exp\bigg(-\frac{i\theta}{4}\big(\sigma^{i,j}_\phi\otimes \mathbb{1} +\mathbb{1}\otimes\sigma^{i,j}_\phi \big)^2\bigg)\,,
\end{equation}
where $\sigma^{i,j}_\phi=\cos(\phi)\sigma_X^{i,j}\pm \sin(\phi)\sigma_Y^{i,j}$ and the indices $i,j$ of the Pauli matrices are related to the levels of the qudits that are being rotated.

In the computational basis, the angular momentum operators are defined as
\begin{align}
    &L_z\ket{m} = m\ket{m},\\
    &L_+\ket{m} = \sqrt{(l-m)(l+m+1)}\ket{m+1},\\
    &L_-\ket{m} = \sqrt{(l+m)(l+m+1)}\ket{m-1},\\
    &L_x = \frac{1}{2}(L_-+L_+),  
\end{align}
where $l=(d-1)/2$ is the index of the angular momentum representation. 
We use these operators to describe our Hamiltonians in the rest of the work. Each unitary operator generated by this set can equally be generated via the universal gate set defined by Eqs.~\eqref{eq: rotation} and \eqref{eq: MS}.

\subsection{Qudit-based variational quantum algorithms}
In VQAs~\cite{cerezo2021variational}, the ground state of a problem Hamiltonian $H_P$ is obtained by utilizing both quantum and classical resources. The quantum part evolves an initial state $\ket{\psi_0}$ to a final state $\ket{\psi_f}$ by means of a parameterized ansatz $U(\boldsymbol{\theta})$. For the initial state, it is necessary to have an overlap with the target state, and for this, usually, the equal superposition state is prepared. The classical part optimizes the set of parameters $\boldsymbol{\theta}$ such that $\ket{\psi_f}$ is close to the desired ground state. This task is accomplished by minimizing a cost function such as the expectation value of $H_P$, the fidelity with the ground state, etc. The performance of a VQA heavily depends upon the design of $U(\vec \theta)$ and the cost function employed~\cite{TILLY20221}.

Among the several ways to define the ansatz, QAOA is a paradigmatic example~\cite{farhi2014quantum}. QAOA takes inspiration from adiabatic quantum computing, where we have a Hamiltonian 
\begin{equation}\label{eq:adiabatic}
    H_a(t) = (1-\lambda(t))H_0 + \lambda(t) H_P,
\end{equation}
where $\lambda(t)$ is the scheduling function interpolating from $H_0$ to $H_P$. The system is initialized in the ground state of $H_0$ and, when the adiabatic condition is satisfied, we can find the ground state of $H_P$ with a high success probability. In QAOA, one splits the continuous time evolution into discrete layers of alternating evolutions under $H_0$ and $H_P$ and replaces the coefficients of both Hamiltonians with trainable parameters $\boldsymbol{\theta}_1$ and $\boldsymbol{\theta}_2$. Therefore, the evolution $U(\boldsymbol{\theta})$ has the form 
\begin{equation}
    U(\boldsymbol{\theta}) =  \prod_{p} e^{-i \theta_{1,p} H_0} e^{-i \theta_{2,p} H_P}\,.
\end{equation}

To minimize the cost function, the classical routines search for optimal parameters from the information obtained by computing its gradients~\cite{Yudai2021normalized}, or by using other heuristics~\cite{Kulshrestha2023}. In most cases, one requires several circuit evaluations depending on the classical optimization method to calculate the parameters for the next step. For example, one requires at least one circuit evaluation per parameter to find gradients for the next step when using gradient-based optimization, yielding a parameter-based measurement overhead. Since the quantum resources in the NISQ era are limited, one has to design the ansatz with a minimum circuit depth and an optimal number of parameters $\lvert\boldsymbol{\theta}\rvert$ to balance the expressivity and measurement overhead.   

Most of the previous works on VQAs have been conducted on qubit systems. Employing qudits is a natural extension, with the major advantage of opening up the possibility of finding more efficient encodings. In particular, qudit systems can enhance the performance of VQA when applied to problems that naturally present multi-dimensional elements, such as for lattice-gauge theories~\cite{deller2023quantum}, integer optimization problems~\cite{Popov2024}, or multi-class classification tasks~\cite{mandilara2023classification}. However, similar to qubits, this approach faces the problem of designing ans\"{a}tze with enough expressivity, while avoiding over-parameterization~\cite{you2022convergence} and the barren plateaus~\cite{ragone2023unified}.

\subsection{Digitized counterdiabatic quantum optimization}
A key issue in adiabatic quantum algorithms for ground-state search is to avoid diabatic transitions to higher excited states. 
One way to mitigate this is { CD} driving, where the Hamiltonian $H_a(t)$ in Eq.~\eqref{eq:adiabatic} is augmented to $H(t)$ given by
\begin{equation}
    H(t) =  H_a(t) + \dot{\lambda}(t) A_\lambda^{(\ell)}, \label{eq: CDham}
\end{equation}
where $A_\lambda^{(\ell)} = \sum_{k=1}^\ell \alpha_{k}(t) Q_{2k-1}$ denotes the auxiliary CD term with $Q_k = [H_a(t),Q_{k-1}]$ and $Q_0 = \partial_{\lambda} H_a(t)$. Here, $\ell$ denotes the order of approximation. { The term $\dot{\lambda}(t) A_\lambda^{(\ell)}$ introduces the CD correction to suppress non-adiabatic transitions. 
} 

The coefficients $\alpha_k(t)$ can be obtained in several ways including action minimization~\cite{doi:10.1073/pnas.1619826114,PhysRevLett.123.090602}, variational optimization~\cite{doi:10.1098/rsta.2021.0282}, machine learning \cite{YaoLinBukov21,Ferrer-Sánchez_2024}, Krylov subspace methods~\cite{PhysRevX.14.011032}, etc. In this work, we consider the action minimization method, where we have to find
\begin{equation}\label{eq:actionmin}
    \min_{\alpha_k} \text{Tr}(G_\ell^2)
\end{equation}
with $G_\ell = Q_0 - i[H_a(t), A_\lambda^{(\ell)}]$. The structure of the operator $G_\ell$ contains a term, $\partial_\lambda H_0$, that is independent of the weights of CD driving, and a sum of parameters that depend on the weights. Thus, we can rewrite this operator as $G_\ell = B_0 + \sum_k \alpha_k B_k$. Note that for first-order CD terms ($\ell = 1$), the operators $B_k$ will contain at most two-body terms when $H_P$ is two-local and $H_0$ is local. We will work with this setup hereafter. 

Once we find the optimal CD terms, we discretize the evolution using the first-order Trotter decomposition given by,
\begin{equation}
    \begin{split}
        U_{\mathrm{Trott}}(0,T) &= \prod_{j=1}^n \prod_{m=1}^M \exp \left[-i \Delta t C_m(j \Delta t) H^{(m)}\right] \\
        &\quad + \mathcal{O}(\Delta t^2) ,
    \end{split}
\end{equation}
where, $n$ is the total number of Trotter steps, i.e., $T = n \Delta t$, $H^{(m)}$ is each of the terms of $H(t)$ and $M$ is the total number of terms present in $H(t)$. One can also utilize a hybrid quantum-classical algorithm known as digitized counterdiabatic QAOA (DCQAOA), where we set coefficients of each term in Eq.~\eqref{eq: CDham} as a trainable parameter optimized by classical optimization~\cite{Pranav2022}. Therefore, one can implement additional unitary $U_{\rm{CD}} = e^{-i\sum_k\theta_k P_k}$ corresponding to the CD term. Here, $P_k$ shows the $k^{\rm{th}}$ term in the linear combination of Pauli operator tensor products corresponding to $A_\lambda^{(\ell)}$. The evolution $U(\boldsymbol{\theta})$ is of the form 
\begin{equation}
    U(\boldsymbol{\theta}) = \prod_p e^{-i \theta_{1,p} H_0} e^{-i \theta_{2,p} H_P}  e^{-i\sum_{k=1}\theta_{2+k,p} P_k}.
\end{equation}

In the impulse regime where $|~\alpha_k(t) \dot \lambda(t)~|\gg |~\lambda(t)~|$, we can neglect the adiabatic Hamiltonian $H_a(t)$ and implement evolutions corresponding to only CD terms~\cite{Pranav2023,cadavid2023efficient}. The unitary evolution for this algorithm is then given by
\begin{equation}
     U(\boldsymbol{\theta}) = e^{-i\sum_{k=1}\theta_k P_k}\,.
\end{equation}

Both DCQAOA and CD-inspired ans\"{a}tze share a parameterization scaling of $\mathcal{O}(N^2)$. For many optimizers, the circuit evaluations needed for each iteration depend upon the number of parameters. To address this challenge, in this work, we leverage symmetries to strategically reduce the number of trainable parameters in a problem-specific manner. 

\section{ Symmetry-enhanced CD-inspired ansatz }\label{sec: Theory symmetric}

Symmetries in physics can be employed to condense the information of a system in the least amount of parameters. This is a property that is used to improve algorithms treating systems consisting of many particles, for instance in the Hartree-Fock method~\cite{Szabo1996}, or DMRG~\cite{White1992}. In this section, we show how to exploit the symmetries of the system to reduce the number of parameters in the variational approach with CD terms.

Let us begin with the symmetries that act on the system, and in particular, on the operator $G_\ell$. Here and in the rest of this article, we focus only on the spatial symmetries of the system~\cite{Zeier2011}. 
These symmetries $\pi$ act on a Hermitian operator $O$ permuting the positions of the qudits such that $\pi O\pi^{-1}=O$, i.e., the action of the symmetry changes the labels of the qudits but leaves the operator invariant. 

\begin{figure}
    \centering
    \includegraphics[width=\linewidth]{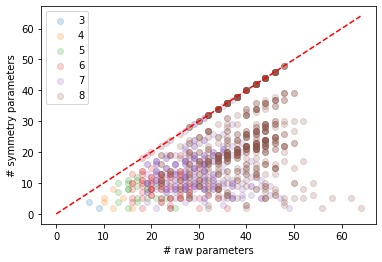}
    \caption{Number of reduced parameters employing spatial permutation symmetries vs.\ total number of parameters without taking into account the symmetries. Here, we have checked all $ZZ$-Ising Hamiltonians as in Eq.~\eqref{eq: ZZIsingQudit} with a connected graph structure up to 8 nodes that can be found in the database from \cite{Coolsaet2023}.}
    \label{fig:reductionParams}
\end{figure}

We denote the set of symmetries of $H_P$ with $S_P=\{\pi_q\}$, where $q$ labels the elements of the set. It is straightforward to see that since $H_0$ is symmetric under the action of all the operations of the permutation group $S_n$, it will also have the ones of $S_P$, $S_P\subseteq S_n$. From this, it follows that the adiabatic Hamiltonian will have the same symmetries as $H_P$. Additionally, the CD Hamiltonian will inherit these symmetries, as the term $\partial_\lambda H(t)$ maintains them as well, $\partial_\lambda H(t)=\pi_q\partial_\lambda H(t)\pi_q^{-1}=\partial_\lambda (\pi_q H(t)\pi_q^{-1})$.
Considering $\{A_k\}$ as the CD terms of the system, the action of each of the symmetries from $S_P$ will only change the labels of these terms ($k\rightarrow \pi_q(k)=k_q$)
\begin{equation}
    \begin{split}
        A_\lambda^{(\ell)} &= \sum_k \alpha_k A_k = \pi_q \left(\sum_k \alpha_k A_k\right) \pi_q^{-1} \\
        &= \sum_{k_q} \alpha_{k_q} A_{k_q}.
    \end{split}
\end{equation}
Now, we can identify each of the terms, such that the only difference after applying the symmetries is in the labels for the parameters. Thus, for any symmetry $\pi_q$ we have
\begin{equation}
    A_\lambda^{(\ell)}=\pi_q A_\lambda^{(\ell)}\pi_q^{-1}=\sum_{k} \alpha_{k_q} A_{k}.
\end{equation}
Now, we formulate the main property of the symmetries. Informally, we say a set of parameters $\{\alpha_k\}_g$ can be grouped, i.e., have the same value, if their labels $k$ form a closed set under the action of all symmetries in $S_P$. Rigorously, we define the set of grouped parameters $\{\alpha_k\}_g$ as the set that fulfills $\pi_q(\alpha_k\in \{\alpha_k\}_g)\in \{\alpha_k\}_g$ $\forall \pi_q\in S_P$.

Interestingly for our problem, this allows us to reduce the total number of variational parameters. Recall the expression of operator $G_\ell$, in which we can identify each of the operators $B_k$ with a CD term $A_k$. The operators $B_k$ inherit the symmetries of $A_k$ as both the terms in $B_k\sim[H_0,A_k]$ share the same symmetries. Thus, the grouping of parameters effectively reduces the number of parameters in the action minimization problem by $\lvert\{\alpha_k\}_g\rvert-1$ for each of the parameter groups. 

To check this reduction, we have numerically computed a particular problem, the ZZ-Ising Hamiltonian $H_P=\sum_i\sigma^z_i+\sum_{i,j\in\mathcal{G}}\sigma^z_i\sigma^z_j$ with a local mixer Hamiltonian $H_0=\sum_i\sigma^x_i$.
As shown in Fig.~\ref{fig:reductionParams}, we have computed the reduction of parameters for all non-trivial problems up to 8 qubits and observed a reduction to the $59\pm19\%$ of the original number of parameters. Additionally, the standard approach and the symmetry-enhanced method used to approximate the CD terms both converge to the same result. Therefore, this reduction of parameters can also be extended to the CD-inspired VQA. Detailed calculations are given in App.~\ref{app:Theory}.

\begin{figure}
    \centering
    \includegraphics[width=0.7\linewidth]{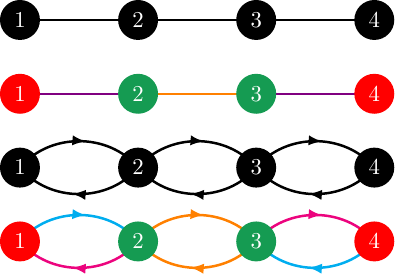}
    \caption{Example of graphs and their symmetries. From top to bottom: $\mathcal{G}$, $\mathcal{G}$ with vertex and edge orbits, $\mathcal{H}$, and $\mathcal{H}$ with the grouped terms in colors, which indeed are the vertex and arc orbits.}
    \label{fig:CrappyGraph}
\end{figure}

To give a clearer idea of the parameter reduction procedure, let us begin with an example of a $N=4$ qudit system in which the problem Hamiltonian is given by
\begin{equation}\label{eq: ZZIsingQudit}
    H_P=\sum_{i=1}^4 {L_z}_i + \sum_{i=1}^3 {L_z}_i{L_z}_{i+1}\,,
\end{equation}
i.e., a Hamiltonian with connectivity graph in the form of a linear path, which we denote as $\mathcal{G}=(V_\mathcal{G}=[1,2,3,4],E_\mathcal{G}=[(1,2),(2,3),(3,4)])$, with $V_\mathcal{G}$ and $E_\mathcal{G}$ the vertices and edges of $\mathcal{G}$, respectively. Given a mixer Hamiltonian of the form $H_0=\sum_{i=1}^4 {L_x}_i$, the CD Hamiltonian is given by
\begin{equation}\label{eq: standard CDs}
    \begin{split}
        A_\lambda^{(1)} &= \sum_{i\in V_\mathcal{G}} \alpha_i {L_y}_i \\
        &\quad + \sum_{\{i,j\}\in E_\mathcal{G}} \left(\alpha_{i,j} {L_y}_i {L_z}_j + \alpha_{j,i} {L_y}_j {L_z}_i\right).
    \end{split}
\end{equation}
Notably, even though the interactions of the problem Hamiltonian do not have any sense of directionality, the interactions in the CD Hamiltonian do. In particular, we cannot interchange ${L_y}_j{L_z}_i\leftrightarrow {L_y}_i{L_z}_j$ without swapping their corresponding parameters. Then, we can identify the terms in the CD Hamiltonian with a directed graph $\mathcal{H}$ in which each edge $E_\mathcal{G}$ is included as an arc of $A_\mathcal{H}$ in both directions as depicted in Fig.~\ref{fig:CrappyGraph}. As a result, we can rewrite the CD Hamiltonian as
\begin{equation}
    A_\lambda^{(1)}=\sum_{i\in V_H} \alpha_i{L_y}_i + \sum_{\{i,j\}\in A_H} \alpha_{i,j} {L_y}_i{L_z}_j.
\end{equation}

To reduce the number of parameters employing this technique, one has to identify the different groups. Above, we defined the grouping of parameters such that it coincides with the definition of the graph orbits~\cite{Siemons1983,Sauvage2024}. For dealing with the cases in which the CD terms have directionality, we use the extended notion of edge orbits to the arcs of a directed graph~\cite{Foldes1980}. For calculating arc orbits, we obtain the edge orbits of the undirected graphs by employing the same algorithm as proposed in Ref.~\cite{Sauvage2024}. We detail the steps to group the parameters in Alg.~\ref{alg:ArcOrbits}. For differentiating the edge orbits on directed graphs from the ones on undirected graphs, we will call the former arc orbits.  

\begin{algorithm}[H]
    \caption{Algorithm for obtaining the arc orbits of the directed graph $\mathcal{H}$}
    \label{alg:ArcOrbits}
    \begin{algorithmic}[H]
        \State Get the undirected graph $\mathcal{G}$ from $\mathcal{H}$
        \State Calculate vertex and edge orbits of $\mathcal{G}$
        \State Define arc orbits as tuple (\textit{arcs}, \textit{tail vertex orbit}, \textit{head vertex orbit}, \textit{edge orbit})
        \ForEach{edge orbit}
            \State Identify the arcs \textit{(h,t)} of $\mathcal{H}$ that fulfill \textit{(h,t)} $=$ \textit{(i,j)} or \textit{(h,t)} $=$ \textit{(j,i)}, where \textit{(i,j)} are the edges of $\mathcal{G}$
            \ForEach{arc}
                \State Identify the vertex orbit of the tail and head
                \State Add the arc to the arc orbit with the same tail vertex, head vertex, and edge orbits
            \EndFor
        \EndFor
    \end{algorithmic}
\end{algorithm}

In our particular case, the vertex orbits of $\mathcal{G}$ are $[1,4]$ and $[2,3]$, and the edge orbits $[(1,2),(3,4)]$ and $[(2,3)]$. Utilizing the Alg.~\ref{alg:ArcOrbits}, we then identify the arc orbits $[(1,2),(4,3)]$, $[(2,1),(3,4)]$ and $[(2,3),(3,2)]$. Subsequently, we can reduce the total number of variational parameters in the system from 10 to 5, as from the symmetries it follows that $\alpha_1=\alpha_4$, $\alpha_2=\alpha_3$, $\alpha_{12}=\alpha_{43}$, $\alpha_{21}=\alpha_{34}$, $\alpha_{23}=\alpha_{32}$.

To sum up, the procedure to generate a symmetry-enhanced CD-inspired ansatz is as follows. First, obtain the CD-terms $A_\lambda^{(\ell)}$ from $H_P$ and $H_0$. Then, compute the arc and vertex orbits of the associated graph. Finally, when parameterizing the ansatz, include the information about the parameter groups obtained in the previous step.
{ It is important to note that in tensor networks and DMRG, leveraging symmetries to reduce parameters may sometimes worsen the convergence~\cite{Hubig2015, hubig2017symmetry}. However, in quantum systems, utilizing symmetries consistently decreases the number of measurements required which is advantageous for near-term devices.} In the next section, we illustrate with numerical examples that this reduction can lead to concrete improvements of key figures of merit of quantum optimization.

\section{Numerical results}\label{sec: benchmark}

To test the validity of the proposed techniques for practical applications in quantum optimization, we need problems whose encoding in qudits is both efficient in terms of the number of particles and mapping of the cost function into a single- and two-body interaction Hamiltonian. Also, we require problems with naturally occurring symmetries. This is the case for some problems defined over lattices or graphs.

\subsection{\textsc{Max}-$3$-\textsc{Cut}}

As a paradigmatic application of quantum optimization, we investigate a combinatorial optimization problem. In this regard, we are not particularly interested in optimization problems in which each instance is heavily determined by weights. Thus, we select the \textsc{Max}-$k$-\textsc{Cut} problem~\cite{Christos1991}, with instances for which no trivial or greedy solution exists.
The objective of this problem is to find the best partition of the vertices into $k$ colors, such that the number of edges connecting two vertices belonging to different colors is maximized. The objective is expressed as
\begin{equation}
    \max_{\mathbf{c}} \sum_{\{i,j\}\in E} \begin{cases}
        1 &\text{if}\ c_i\neq c_j,\\
        0 &\text{else},
    \end{cases}
\end{equation}
where $\mathbf{c}=\{c_i\}^N$ is the vector assigning each vertex $i$ to the colour $1\leq c_i\leq k$ and the size of the problem $N$ is the number of the vertices of the graph. Since the instances are uniquely defined by a graph, the symmetries from the graph will be used to reduce the number of parameters in our variational approach. The conversion of this classical problem into the problem of finding the ground state of a Hamiltonian can be done with two-local even-order terms. For the particular case of the \textsc{Max}-$3$-\textsc{Cut}, we have
\begin{equation}
    H_P = \sum_{\{i,j\}\in E} {L_z}_i{L_z}_j -2({L_z}_i^2+{L_z}_j^2) + 3{L_z}_i^2{L_z}_j^2.
\end{equation}
Even though we limit our simulations to the \textsc{Max}-$3$-\textsc{Cut}, the results here are easily extensible to any $k$, since to obtain the Hamiltonian for the general \textsc{Max}-$k$-\textsc{Cut} problem one just needs to solve a linear system of $\mathcal{O}(k^2)$ equations, as shown in App.~\ref{app:maxkcut}.

\begin{figure}[t]
    \centering
    \includegraphics[width=\linewidth]{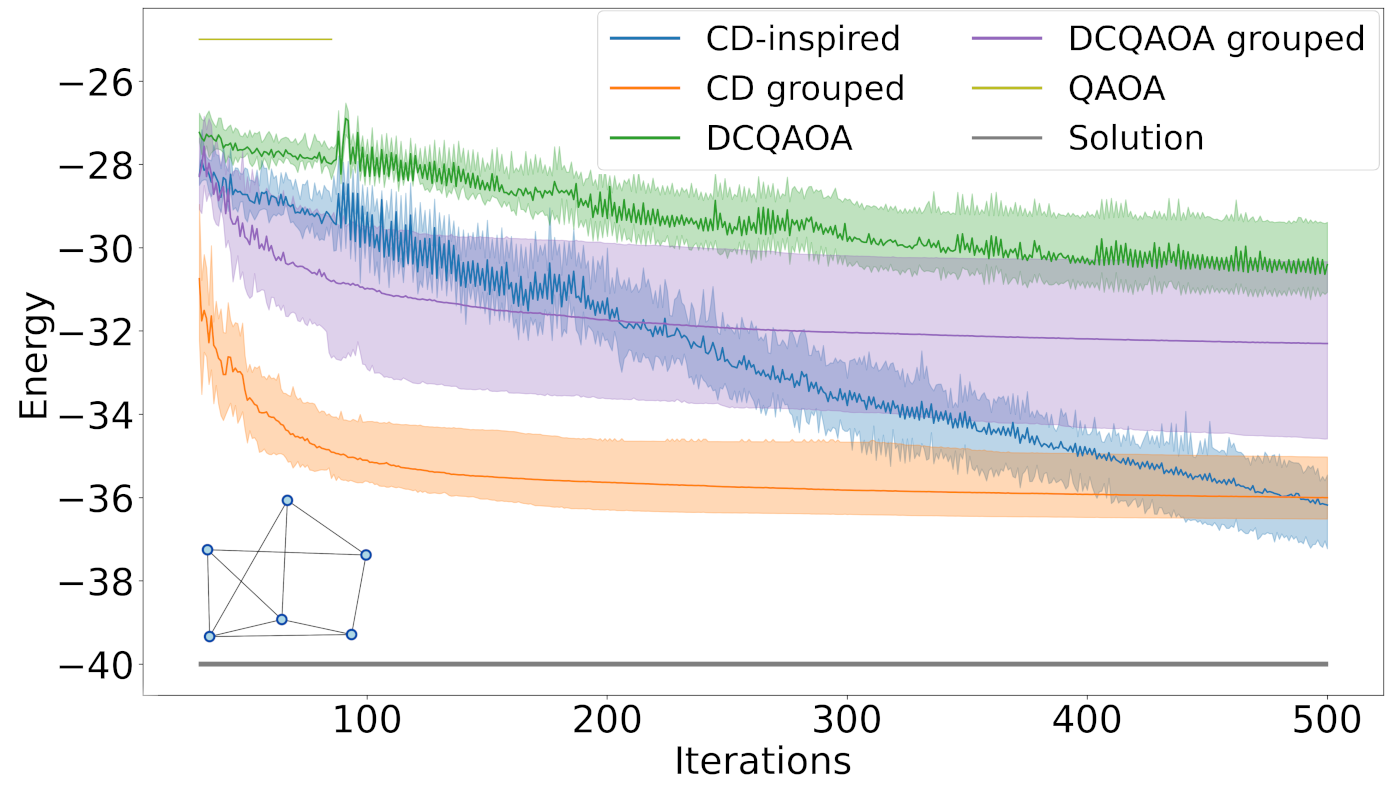}
    \caption{Expected value of the energy at the end of each iteration of the variational procedure for a Max-3-Cut instance. We show the results for the different ans\"{a}tze for a single layer. We only show the results after 30 steps of the classical optimizer to discard the initial fluctuations. As shown, the CD ansatz with the grouped parameters shows a faster convergence compared to the fully parameterized CD ansatz. After a high number of iterations, the fully parameterized ansatz obtains better results due to its higher expressivity. These results are also true for the DCQAOA variants. QAOA fails to solve the problem due to its limited expressivity even though the depths of the ans\"{a}tze are similar.}
    \label{fig:singleResults}
\end{figure}

Using this problem, we have tested the performance of our proposal with two $p=1$ layered ans\"{a}tze against several similar approaches and the QAOA. We start with the DCQAOA ans\"{a}tze at first order as proposed in Ref.~\cite{Pranav2022}. Using our approach, we refine this ans\"{a}tze by grouping the parameters of the CD terms. We also tested similar circuits in which we only use the CD terms, which are also typically known as CD-inspired ans\"{a}tze~\cite{Pranav2023}. As a baseline for comparison, we employed the usual QAOA for qudits as proposed originally in Refs.~\cite{farhi2014quantum, deller2023quantum}. For the benchmark, we have selected 4 instances of the smallest non-trivial size with $N=6$ qudits of $d=3$ levels, commonly known as qutrits. The graphs for the instances have been selected from the database in Ref.~\cite{Coolsaet2023}, with the identifiers HoG Id 728, 220, 730, and 758. The selection criteria we employed are graphs with 6 nodes, fully connected and non-planar~\cite{Hadlock1975}. We have run up to 500 iterations of the COBYLA optimizer~\cite{Powell1994} for each of the classical update loops because this algorithm does not require computing costly derivatives. For each problem instance, we have run multiple different initializations with uniformly random parameters. 

The results we have obtained for the expected value of the energy for the first graph are shown in Fig.~\ref{fig:singleResults}. The full comparison is given in App.~\ref{app:maxkcut}. Additionally, approximation ratios are shown in Tab.~\ref{tab:my-table}, which are defined as $\mathcal{R}=\braket{H_P}/E_0$ with $E_0$ the ground state energy. We observe that all four algorithms implementing CD-terms reach lower energies than QAOA with few iterations. Moreover, we see that the ans\"{a}tze with grouped terms quickly reaches energies close to the convergence energy. For both the grouped and non-grouped versions, the DCQAOA ans\"{a}tze seems to need a higher number of iterations to converge. This is counter-intuitive as the expressibility of the DCQAOA ans\"{a}tze is the same as that of the CD ans\"{a}tze if we set the QAOA parameters to 0.  However, the reduction of the convergence speed might be due to the appearance of barren plateaus~\cite{ragone2023unified} and over-parametrization effects~\cite{you2022convergence}. We also stress that even though the performance of the grouped ansatz is slightly worse than the ungrouped one, we still achieve a good result with a reduced amount of variational parameters. Although the { simulations have} not been extensive, for all four problem instances checked, the grouping method consistently shows a performance improvement.
Our approach not only improves the convergence speed or the approximation ratio as shown, but more importantly decreases the number of circuit evaluations.

\begin{table}[t]
\caption{Mean approximation ratios and standard deviation for the algorithms at the final iteration for the different problems. Values in bold showcase which the obtained result yields a better $\mathcal{R}$ values than the classical approximation ratio of 0.800217 \cite{Frieze1997}. Underlined values mark the best performance for each problem in terms of the mean approximation ratio $\mathcal{R}$. QAOA converged to the same value of energy for each graph up to machine precision.} 
\label{tab:my-table}
\resizebox{\columnwidth}{!}{%
\begin{tabular}{| p{48pt} |l|l|l|l|}\hline
Problem graph Id & \multicolumn{1}{c|}{728}      & \multicolumn{1}{c|}{220}      & \multicolumn{1}{c|}{730}      & \multicolumn{1}{c|}{758}      \\\hline
CD all           & {\ul\textbf{0.90 ± 0.04}}     & {\ul \textbf{0.96 ± 0.01}}    & \textbf{0.85 ± 0.03}          & \textbf{0.91 ± 0.03}       \\\hline
CD grouped       & {\ul\textbf{0.90 ± 0.05}}      & \textbf{0.95 ± 0.03}          & {\ul \textbf{0.92 ± 0.06}}    & {\ul \textbf{0.92 ± 0.03}} \\\hline
DCQAOA all       & 0.76 ± 0.03                   & \textbf{0.91 ± 0.04}          & 0.74 ± 0.04                   & \textbf{0.81 ± 0.01}       \\\hline
DCQAOA grouped   & \textbf{0.81 ± 0.06}          & \textbf{0.90 ± 0.04}          & \textbf{0.81 ± 0.04}          & \textbf{0.87 ± 0.06}       \\\hline
QAOA             & 0.63                   & 0.76                   & 0.63                 & 0.69            \\\hline
\end{tabular}%
}
\end{table}
\begin{figure*}[t]
    \centering 
    \includegraphics[width = \linewidth]{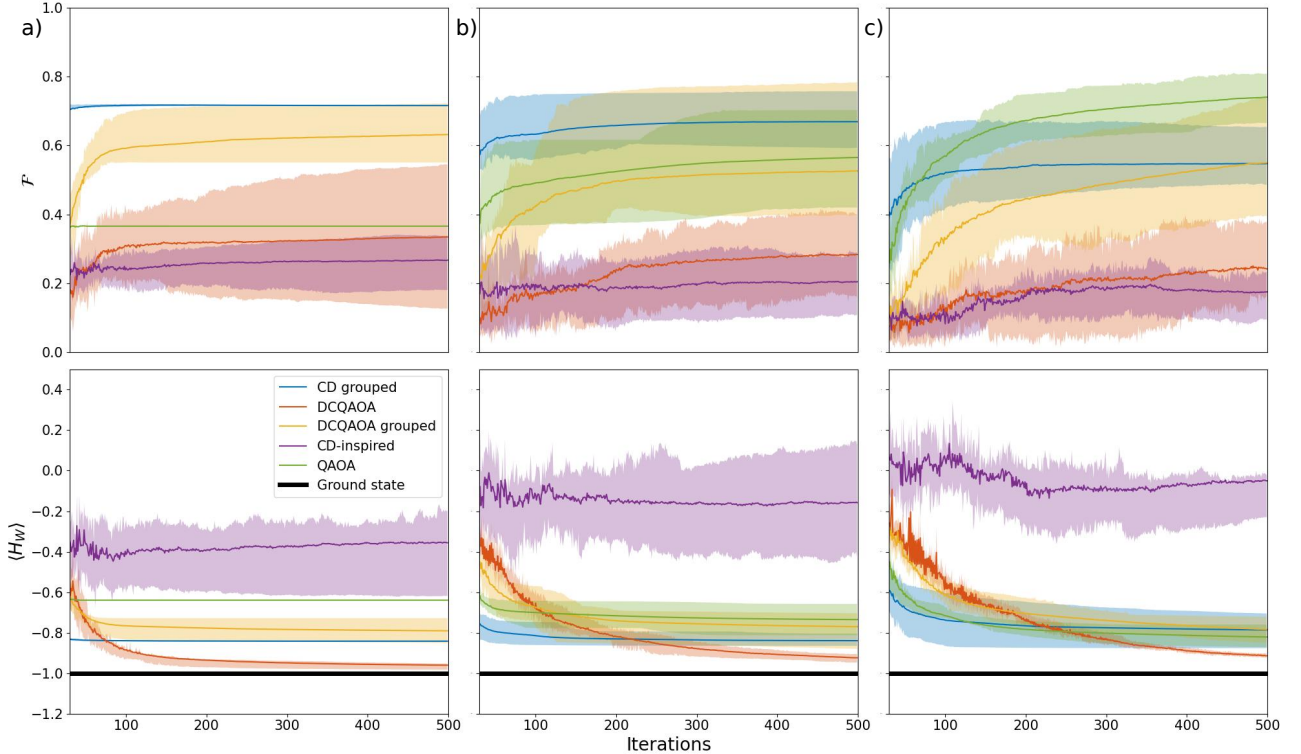}
    \caption{Fidelity defined in Eq.~\eqref{eq: fidlity} (top) and the expectation value of the Hamiltonian of Eq.~\eqref{eq: Hamiltonian W} (bottom) for the ans\"{a}tze considered, shown as a function of the optimizer iteration. The results refer to the case $N=3$ and the three figures show the results for different numbers of layers: (a) $p=2$, (b) $p=3$, and (c) $p=5$, as ordered from left to right. Also here, we omit the first $30$ iterations to avoid showing initial fluctuations of the optimizer. We see that the grouped CD ansatz outperforms the other ones for a low number of layers, as expected.}
    \label{fig: W results1}
\end{figure*}

\subsection{$W$ state preparation}
As a second test, we study the ability  of the CD symmetrized ansatz to prepare the $W$ state of a system of $N$ qutrits:

\begin{equation}\label{eq: dicke qutrit}
    \ket{W} = \frac{1}{\sqrt{N}} \sum_{i=1}^N P_i (\ket{2}^{\otimes N-1}\ket{0}),{}
\end{equation}
where $P_i$ indicates permutations of the labels of the local basis states. Note that the $W$ state is also a qutrit Dicke state $D_k^N$ with $k=1$, as shown in App.~\ref{app: dicke appendix}. This is an equal superposition of all the possible ground states  of the Hamiltonian 
\begin{equation}\label{eq: Hamiltonian W}
    H_P = \left( \frac{N}{2}-1 \right)\sum_{i=1}^{N}{L_z}_i + \frac{1}{2}\sum_{j<i}{L_z}_i{L_z}_j.
\end{equation}
This Hamiltonian can be represented by a fully connected graph with a local potential on each vertex.
Grouping the coefficients, in this case, is trivial since the graph is fully connected and the resulting CD Hamiltonian is Eq.~\eqref{eq: standard CDs} with $\alpha_i=\alpha\hspace{2mm} \forall \hspace{2mm} i,\hspace{2mm}\alpha_{i,j}=\tilde{\alpha}\hspace{2mm} \forall \hspace{2mm} i,j$. 
This grouping of the coefficients enforces the symmetry of the unitary transformation on the system, which would be lost if all the parameters were left independent. 
Since the $W$ state is the equal superposition of the $N$ possible ground states, it is necessary to guide the evolution towards this specific state and not to a generic superposition of ground states. 
The metric with which we measure the efficiency of the state preparation is the fidelity of the final state with the state Eq.~\eqref{eq: dicke qutrit}. Given that the states are pure, the fidelity simply becomes the expectation value of the evolution operation over the initial and target state:
\begin{equation}\label{eq: fidlity}
    \mathcal{F}=|\braket{W|U(\boldsymbol{\theta})|\psi_0}|^2.
\end{equation} 
The cost function that we minimize in the algorithm is the expectation value of $H_P$, thus the optimization procedure will drive the state towards lower energy. In App.~\ref{app: dicke appendix}, we also show the results that are obtained by maximizing the fidelity instead of minimizing the energy.
The symmetry of the $W$ state will be imposed on our proposed ans\"{a}tze. Thus, we expect a better performance in terms of fidelity compared to the fully parametrized ones, even if the latter has final energy closer to the ground state energy.

We executed the algorithm for the ans\"{a}tze described in Sec.~\ref{sec: Theory symmetric} from $p=1$ to $p=5$ layers, $N=3$ qudits, and show the results in Fig.~\ref{fig: W results1}.
The symmetry-enhanced CD ansatz outperforms QAOA for a small number of layers but gets worse as their number increases. This can be attributed to the fact that for low values of $p$ we are in the impulse regime, while for larger $p$ the ansatz is closer to the adiabatic regime. Furthermore, the fully parametrized ans\"{a}tze approximate the ground state energy faster but have worse results in terms of fidelity. 
Specifically, for just a few layers ($p=2$) we can achieve fidelity with the $W$ state of $\mathcal{F}\sim 71\%$.
\begin{figure*}[ht]
    \includegraphics[width=0.49\linewidth]{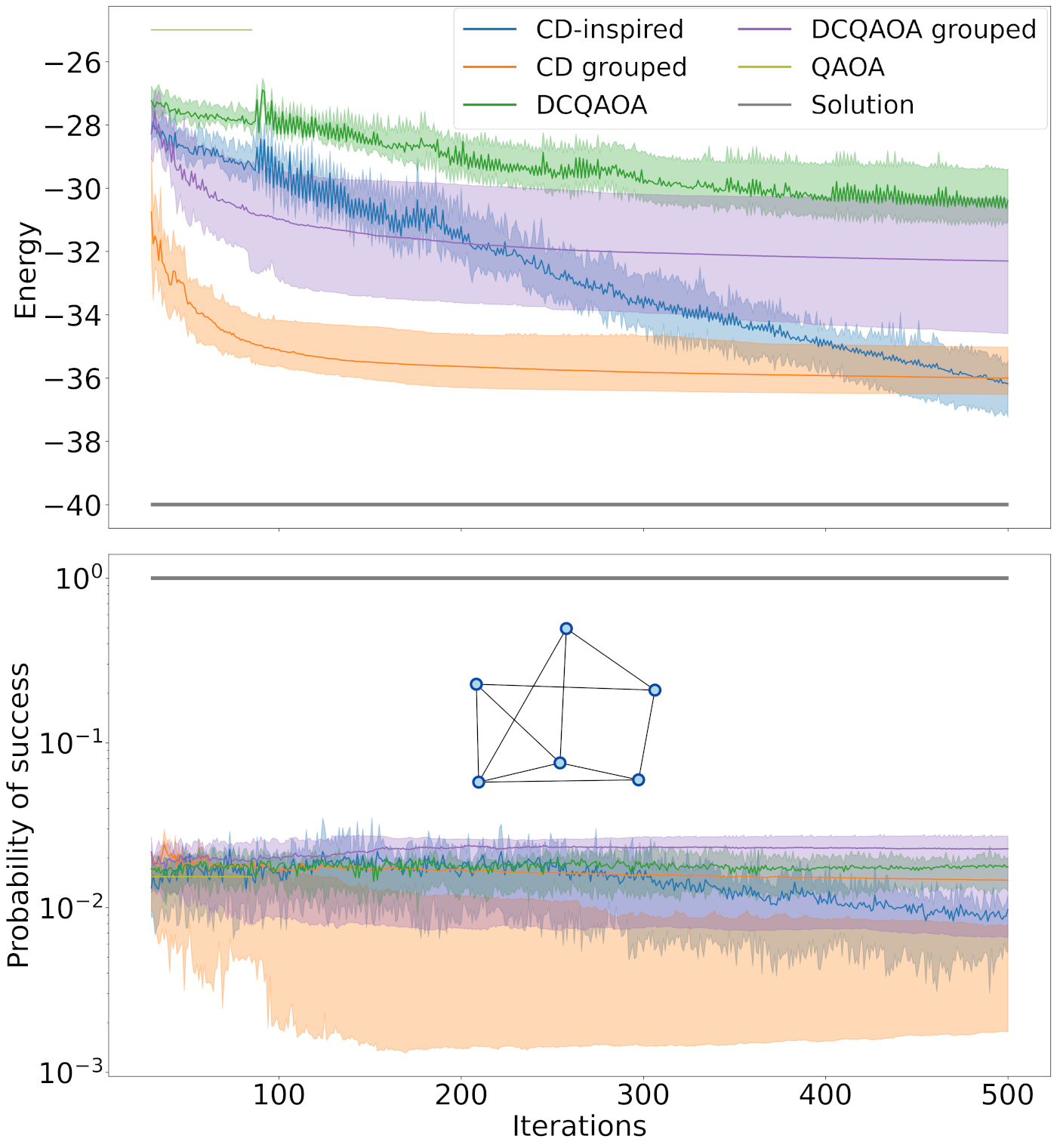}
    \includegraphics[width=0.49\linewidth]{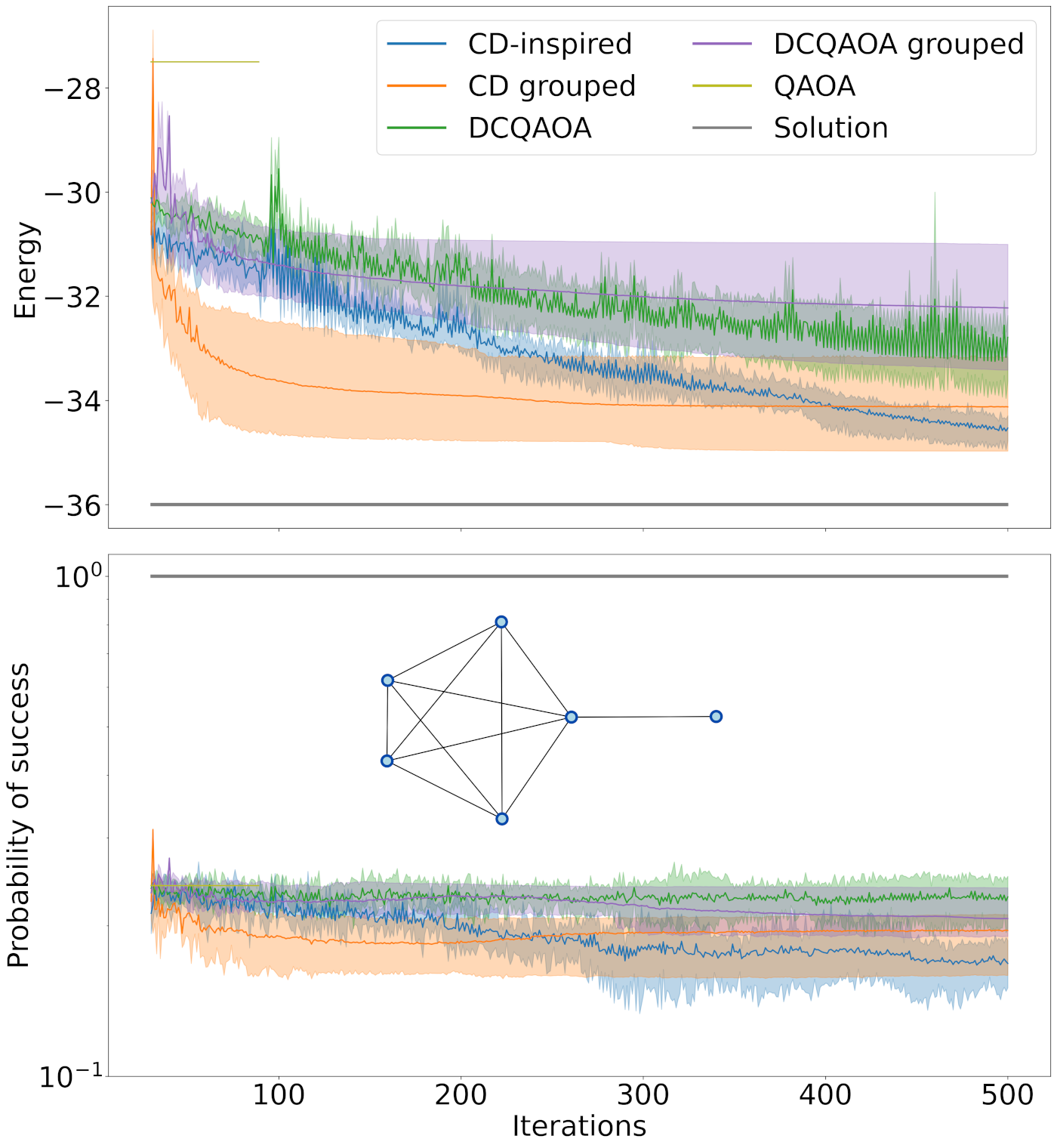}\\
    \includegraphics[width=0.49\linewidth]{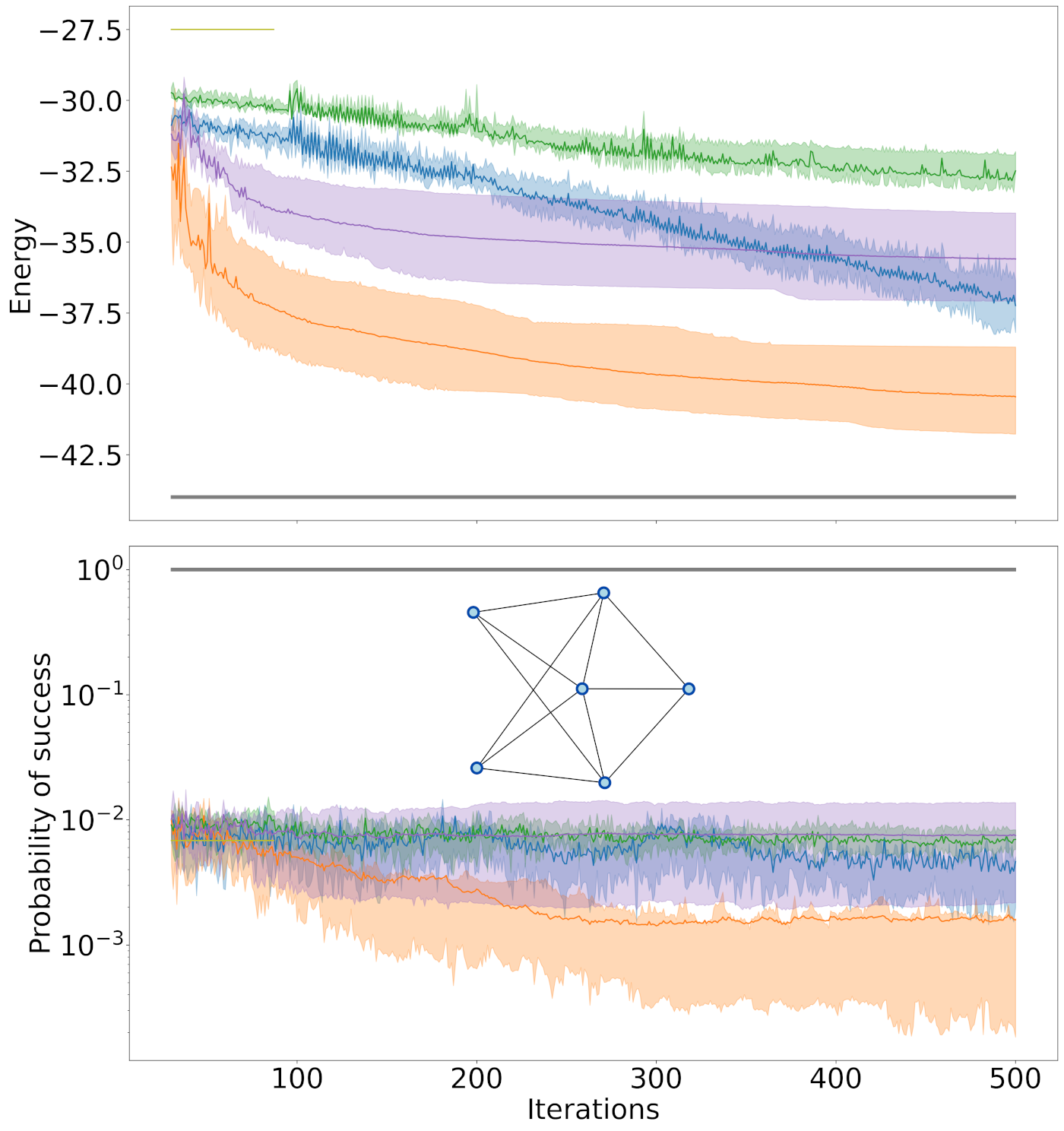}
    \includegraphics[width=0.49\linewidth]{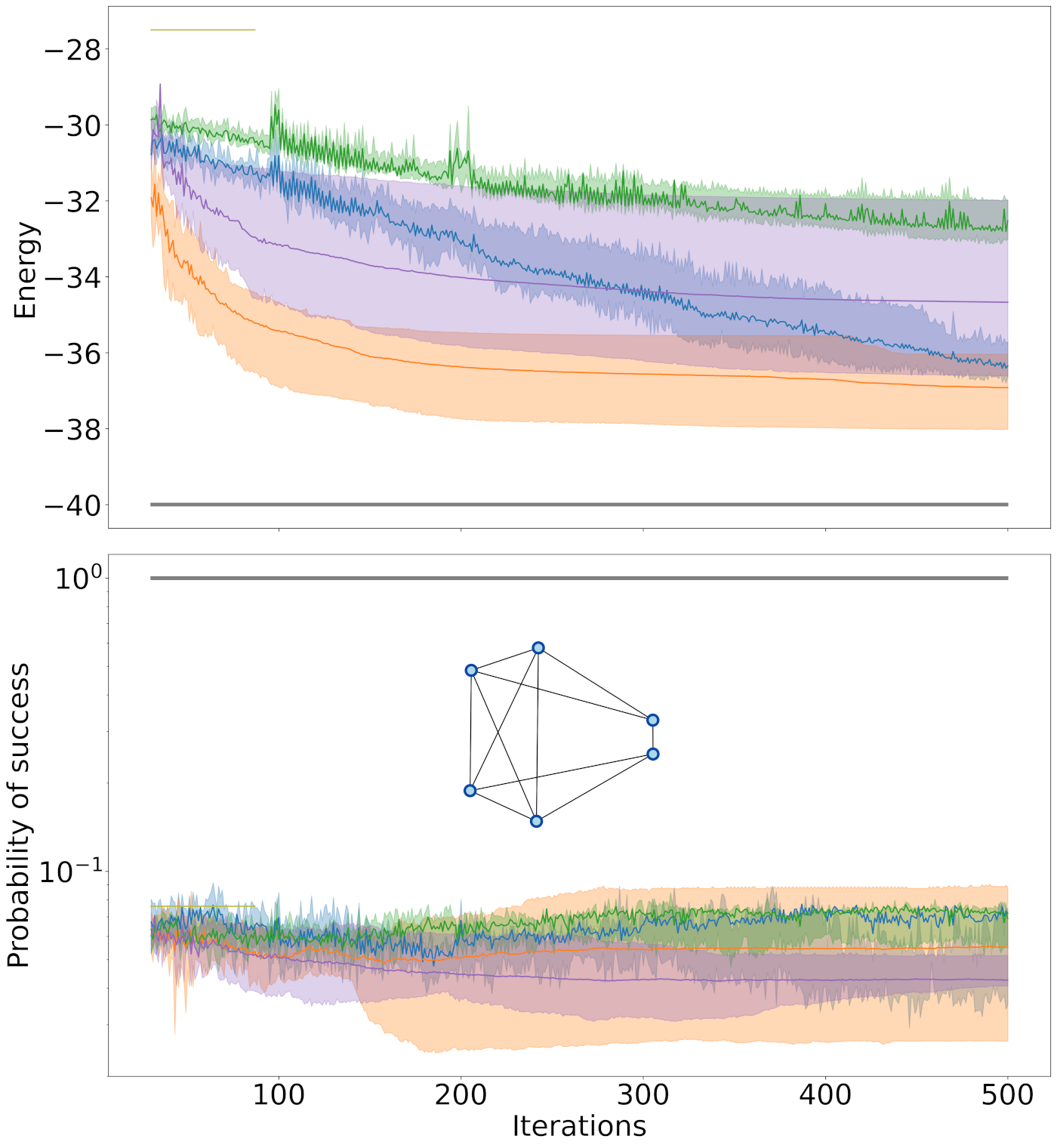}
    \caption{Results for the different algorithms and instances for the Max-3-Cut. The solid lines correspond to mean values and the colored area corresponds to the interquartile range. The upper plots show the energy at each iteration step. The plots in the middle of each sub-figure show the probability of measuring any of the states that correspond to the correct solution. The inset graphs correspond to each of the graphs that define the instances. We have run 20, 12, 12, and 10 runs respectively for each problem using uniformly random initial parameters.}
    \label{fig:Max3CutRuns}
\end{figure*}

\section{Conclusion}\label{sec: conclusion}
In this article, we have introduced a symmetry-enhanced CD-inspired ansatz for variational qudit algorithms. The main advantage of incorporating CD terms lies in the digital circuit compression 
compared to adiabatic protocols. To further speed up the convergence rate of the algorithms, we have utilized the symmetries of the problem to group the variational parameters. The reduction of the parameters also decreases the measurement overhead needed to advance during the optimization process. This is achieved by analyzing how the symmetries are propagated from the problem Hamiltonian to the adiabatic gauge potential and finally to the CD-ans\"{a}tze. 

To test the validity of our approach, we have addressed two problems: the classical optimization problem \textsc{Max}-$k$-\textsc{Cut} and the preparation of a $W$ state using $p=2,3,5$ layer ans\"{a}tze. For the former, we have run a test on four non-trivial instances and compared our approach against various state-of-the-art algorithms. For single-layer ansatz, we have observed better performance of all approaches that employ CD terms. Additionally, when studying the effect of the symmetries, we confirm a faster convergence and better performance. In the case of $W$ state preparation, we have investigated the problem with the same suite of algorithms, demonstrating the utility of our strategy in preparing this state with fewer resources.

The compression in circuit depth, information encoding, and parameters positions this approach as a feasible candidate for experimental implementations. Further works could also discuss if using approximate symmetries could be beneficial, as it would open more possibilities for reducing variational parameters with them in a wider set of problems. Another possibility would be to explore the possibilities of efficient qubit to qudits mappings and their impact on the performance of CD protocols.

\section*{Acknowledgements}
Alberto Bottarelli and Mikel Garcia de Andoin contributed equally to this work.
A.B.\ acknowledges funding from the Honda Research Institute Europe.
P.H.\ acknowledges funding by the European Union under Horizon Europe Programme, Grant Agreement 101080086 — NeQST.
This project has received funding from the Italian Ministry of University and Research (MUR) through the FARE grant for the project DAVNE (Grant R20PEX7Y3A), was supported by the Provincia Autonoma di Trento, and Q@TN, the joint lab between University of Trento, FBK—Fondazione Bruno Kessler, INFN—National Institute for Nuclear Physics, and CNR—National Research Council.
Project funded under the National Recovery and Resilience Plan (NRRP), Mission 4 Component 2 Investment 1.4 - Call for tender No. 1031 of 17/06/2022 of Italian Ministry for University and Research funded by the European Union – NextGenerationEU (proj.\ nr.\ CN\_00000013).
Project DYNAMITE QUANTERA2\_00056 funded by the Ministry of University and Research through the ERANET COFUND QuantERA II – 2021 call and co-funded by the European Union (H2020, GA No 101017733). 
M.GdA., P.C., K.P. and M.S. acknowledge founding from OpenSuperQ+100 (Grant No. 101113946) of the EU Flagship on Quantum Technologies,
EU FET-Open project EPIQUS (Grant No. 899368),
Spanish Ramón y Cajal Grant No. RYC-2020-030503-I funded by MCIN/AEI/10.13039/501100011033 and by “ERDF A way of making Europe” and “ERDF Invest in your Future”,
Spanish Ministry for Digital Transformation and of Civil Service of the Spanish Government through the QUANTUM ENIA project call - Quantum Spain,
EU through the Recovery, Transformation and Resilience Plan – NextGenerationEU within the framework of the Digital Spain 2026 Agenda,
Basque Government through Grant No. IT1470-22,
IKUR Strategy under the collaboration agreement between Ikerbasque Foundation and BCAM on behalf of the Department of Education of the Basque Government.
M.GdA. acknowledges support from the UPV/EHU and TECNALIA 2021 PIF contract call, from the Basque Government through the "Plan complementario de comunicación cúantica" (EXP.2022/01341) (A/20220551), from the Basque Government through the ELKARTEK program, project "KUBIT - Kuantikaren Berrikuntzarako Ikasketa Teknologikoa" (KK-2024/00105), and from the Spanish Ministry of Science and Innovation under the Recovery, Transformation and Resilience Plan (CUCO, MIG-20211005), Spanish CDTI through Plan complementario Comunicaci\'on cu\'antica (EXP. 2022/01341)(A/20220551).
Views and opinions expressed are however those of the authors only and do not necessarily reflect those of the European Commission, the European Union or of the Ministry of University and Research. Neither the European Union nor the granting authority can be held responsible for them. 

\appendix

\section{Theory of symmetry grouped CD terms}\label{app:Theory}

Assume that the problem Hamiltonian $H_P$ has some symmetries $S=\{\pi_k\}$ which can belong to the group of permutations. Now, the Hamiltonian is invariant under the action of $S$
\begin{equation}
    H_P=\pi_k H_P \pi_k^{-1}\ \forall k.
\end{equation}
From this, $H_a(t)$ will be also symmetric
\begin{equation}
    H_a(t)=\pi_k H_a(t) \pi_k^{-1}.
\end{equation}

The objective is to speed up the usual annealing process by employing CD terms as detailed in the main text. Recall the expression of the CD term at first order
\begin{equation}
    Q_1(t)=[H(t),\partial_t H(t)].
\end{equation}

The CD Hamiltonian will inherit the symmetries $S$ exactly, since
\begin{equation}
\begin{split}
    Q_1(t)&=\pi_k Q_1(t)\pi_k^{-1}=\pi_k [H_a(t),\partial_t H_a(t)]\pi_k^{-1}\\
    &=[\pi_k H_a(t)\pi_k^{-1},\pi_k\partial_t H_a(t)\pi_k^{-1}]\\
    &=[\pi_k H_a(t)\pi_k^{-1},\partial_t (\pi_k H_a(t)\pi_k^{-1})]\\
    &=[H_a(t),\partial_t H_a(t)].
\end{split}
\end{equation}
The only effect of acting with the permutation symmetries is a change in the labels of the CD terms. To solve the problem of the original labeling of the qubits and the permuted one, one can identify the identical elements associated with the labels of the parameters, and group them.

\textit{Example:} Assume we have a CD Hamiltonian with 7 different terms and a single symmetry $\pi$. When calculating $Q_1$, one gets the following structure,
\begin{equation}
    \begin{split}
        Q_1 &= \alpha_1 B_1 + \alpha_2 B_2 + \alpha_3 B_3 + \alpha_4 B_4 \\
            &\quad + \alpha_5 B_5 + \alpha_6 B_6 + \alpha_7 B_7.
    \end{split}
\end{equation}

If we apply the symmetry and recalculate the Hamiltonian, we get
\begin{equation}
    \begin{split}
        \pi Q_1 \pi^{-1} &= \alpha_3 B_1 + \alpha_2 B_2 + \alpha_1 B_3 + \alpha_7 B_4 \\
                          &\quad + \alpha_6 B_5 + \alpha_5 B_6 + \alpha_4 B_7.
    \end{split}
\end{equation}

As both Hamiltonians are equal, we can identify term by term and reduce the number of parameters. In this case, we see that $\alpha_1=\alpha3$, $\alpha_4=\alpha_7$ and $\alpha_5=\alpha_6$.

\begin{figure*}[t]
    \centering
    \includegraphics[width = \linewidth]{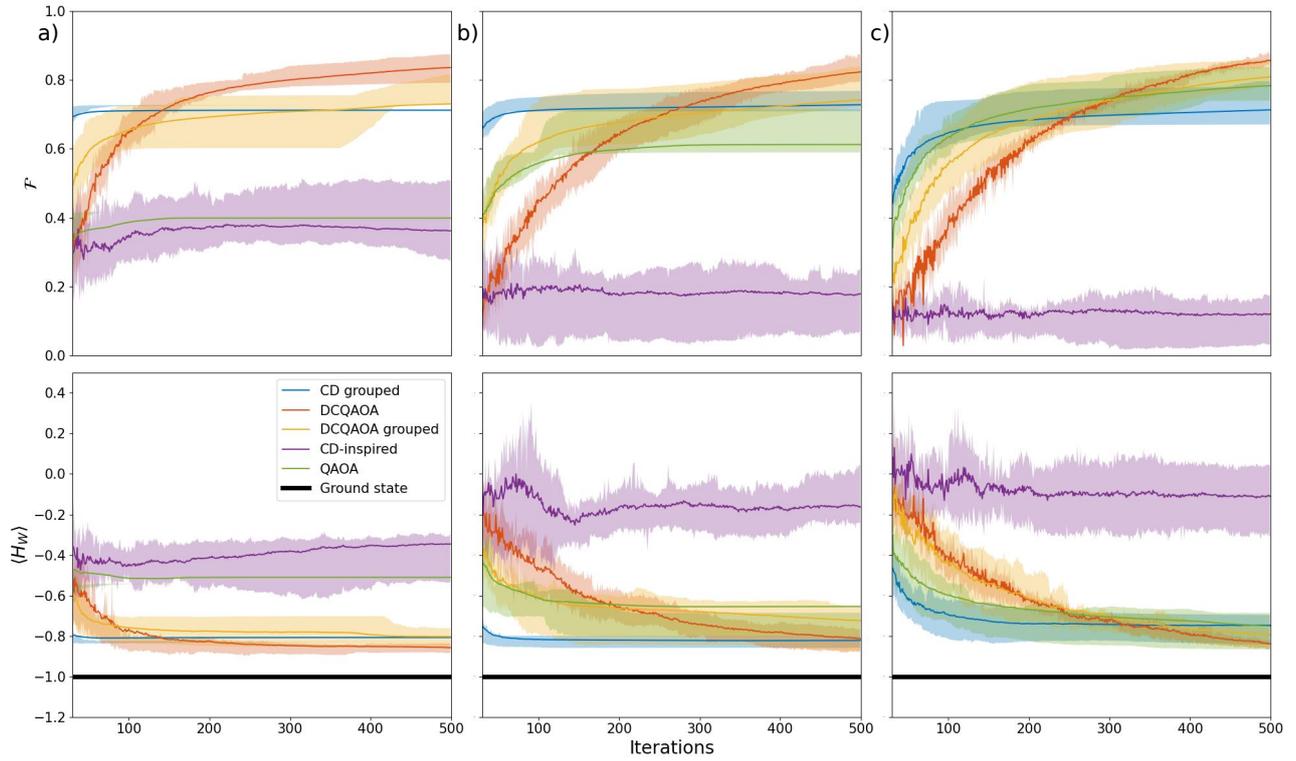}
    \caption{Fidelity and the expectation value of the Hamiltonian of Eq.~\eqref{eq: Hamiltonian W} (bottom) obtained by maximizing the fidelity, shown as a function of the optimizer iteration. Results for $N=3$ and layers: $p=2$(a), $p=3$(b), and $p=5$(c). The first $30$ iterations not shown. }
    \label{fig: fidelity results}
\end{figure*}

\section{Formulation of \textsc{Max}-$k$-\textsc{Cut} on qudits and results for $k=3$ simulations}\label{app:maxkcut}

We want to maximize the number of vertices connecting $k$-different regions in a graph. For the usual \textsc{Max}-$2$-\textsc{Cut}, the elements of the problem Hamiltonian corresponding to each of the edges is
\begin{equation}
    H_{i,j}=4L_{zi}L_{zj}
\end{equation}
in the usual $d=2$ qudits description. The pool of CD terms at first order is
\begin{equation}
    Q_1=\{L_{yi}L_{zj}\}
\end{equation}
For \textsc{Max}-$3$-\textsc{Cut}
\begin{equation}
    H_{i,j}=L_{zi}L_{zj}-2(L_{zi}^2+L_{zj}^2)+3L_{zi}^2L_{zj}^2,
\end{equation}
\begin{equation}
    Q_1=\{{L_y}_i{L_z}_j,L_zL_y+L_yL_z,{L_z}_i{L_y}_j+{L_y}_j{L_z}_i,L_z^2\},
\end{equation}
with $d=3$ qudits. For \textsc{Max}-$4$-\textsc{Cut}
\begin{multline}
    H_{i,j}=\frac{365}{72}L_{zi}L_{zj}-\frac{5}{8}(L_{zi}^2+L_{zj}^2)+\frac{1}{2}L_{zi}^2L_{zj}^2\\
    -\frac{41}{18}(L_{zi}^3L_{zj}+L_{zi}L_{zj}^3)+\frac{10}{9}L_{zi}^3L_{zj}^3
\end{multline}
for $d=4$ qudits.

We see that all terms involved in these problem Hamiltonians contain each an even number of $L_z$ operators such that the maximum degree of this operator in a single qudit is $k-1$. Calculating these coefficients is as difficult as solving a linear system of $\sum_{d=1}^k \lceil d/2\rceil - 1$ equations, which scales as $\mathcal{O}(k^2)$.

In Fig.~\ref{fig:Max3CutRuns} we show the results for the experimentation done solving the \textsc{Max}-$3$-\textsc{Cut} problem for 4 instances generated from non-planar graphs with 6 vertices. We show the results in terms of the energy of the final states at each iteration. This is indeed the cost function we employed for the classical optimization loop. Additionally, we show the probability of measuring the optimal solution for each problem. We observe that even in the worst cases the success probability is at least one order of magnitude higher than random guessing. This behavior is consistent with the fact that we are using the energy as the cost function and not the fidelity.

\section{Dicke states and optimization of fidelity for $W$ state preparation}\label{app: dicke appendix}

Dicke states are important metrological states due to their connection to the theory of quantum communication and multipartite entanglement\cite{bengtsson2016brief}.
For a set of $N$ qudits, a possible choice of the Dicke state can be
\begin{equation}
    D_\kappa^N= \binom{N}{\kappa}^{-\frac{1}{2}} \sum_i P_i (\ket{d-1}^{\otimes N-\kappa}\ket{0}^{\otimes \kappa}),
\end{equation}
where $P_i$ indicates all the possible permutations of basis states. This is a ground state of the Hamiltonian 
\begin{equation}
    H_\kappa^N = \left( \frac{N}{2}-\kappa \right)\sum_{i=1}^{N}{L_z}_i + \frac{1}{2}\sum_{j<i}{L_z}_i{L_z}_j.
\end{equation}
In the previous For $\kappa=1$ and qutrits, the corresponding Hamiltonian is the one of Eq. \eqref{eq: Hamiltonian W}, and the corresponding Dicke states are 
\begin{equation}
    D_1^N= \frac{1}{\sqrt{N}} \sum_i P_i (\ket{2}^{\otimes N-1}\ket{0}),
\end{equation}
also known as the $W$ states.

In the main text, we simulated variational algorithms by minimizing the expectation value of the energy Eq.~\eqref{eq: Hamiltonian W}. Since our goal is to prepare a $W$ state for three qudits, we try a direct approach by using the same variational ans\"{a}tze but maximizing the fidelity in Eq.~\eqref{eq: fidlity}. A couple of considerations must be made in this procedure. This cost function can be represented as $\mathrm{Tr}[U(\boldsymbol{\theta})\ket{\psi_0}\bra{\psi_0}U(\boldsymbol{\theta})^\dagger\ket{W}\bra{W}]$. 
Even though we cannot directly relate the Hamiltonian in Eq.~\eqref{eq: Hamiltonian W}, we can keep parallel with the original problem by considering that the $W$ state of Eq.~\eqref{eq: dicke qutrit} is still a ground state of the original Hamiltonian and thus maximizing the fidelity will also minimize the energy. As shown in the results of Fig.~\ref{fig: fidelity results}, this algorithm obtains better results in approximating the $W$ state ($\mathcal{F}\sim 85\% $ in the best case) and shows how CD terms have better performance than regular QAOA for few layers ($q\leq 4$).


%

\end{document}